\documentclass[12pt]{article}

\usepackage[dvips]{graphicx}
\usepackage{array}
\usepackage{graphicx}
\usepackage{subfigure}
\usepackage{amsmath}
\usepackage[indent,bf]{caption2}
\usepackage{rotating}
\usepackage{setspace}
\usepackage{longtable}
\usepackage{cancel}
\usepackage{fancyheadings}

\newcommand{\be}{\begin{equation}}
\newcommand{\ee}{\end{equation}}
\newcommand{\bea}{\begin{eqnarray}}
\newcommand{\eea}{\end{eqnarray}}
\newcommand{\bean}{\begin{eqnarray*}}
\newcommand{\eean}{\end{eqnarray*}}
\begin{document}

\rhead{CERN-PH-TH/2008-212}
\renewcommand{\headrulewidth}{0pt}

\title{Leptonic Pion Decay And Physics Beyond The Electroweak Standard Model}
\author{Bruce A. Campbell$^{1,2}$ and Ahmed Ismail$^{1,3}$\\ \\
$^1$Department of Physics, Carleton University,\\Ottawa ON K1S 5B6, Canada\\
$^2$Theory Division, PH Department, CERN,\\CH-1211 Geneva 23, Switzerland\\
$^3$Department of Physics, Stanford University,\\Stanford California 94305, U.S.A.}
\maketitle
\thispagestyle{fancy}
%%%%%%%%%%%%%%%%%%%%%%%%%%%%%%%%%%%%%%%%%%%
%\address{$^1$Department of Physics, Carleton University,\\
%Ottawa ON K1S 5B6, Canada \\
%$^2$Division PH-TH, CERN,\\ CH1211 Geneva 23, Switzerland\\$^3$Department 
%of Physics, Stanford University, \\Stanford California 94305, U.S.A.}
%%%%%%%%%%%%%%%%%%%%%%%%%%%%%%%%%%%%%%%%%%%%%%%%%%%%%%%%%%%%%%%%%%%%%%%%%
%%%%%%%%%%%%%%%%%%%%%%%%%%%%%%%%%%%%%%%%%%%%%%%%%%%%%%%%%%%%%%%%%%%%%%%%%
%\date{}
\abstract{The ratio of branching ratios in leptonic pion decay $R_{\pi} \equiv (\Gamma(\pi^- \rightarrow e \nu_e))/(\Gamma(\pi^- \rightarrow \mu \nu_\mu))$ is a powerfully sensitive probe of new interactions beyond the electroweak standard model. This is due to the chirality suppression of the standard model amplitude for the decay, which results in a precise prediction for the ratio, and suppressed amplitudes for new contributions to interfere with. We calculate, including QCD corrections, the contributions to $R_{\pi}$ arising from a broad selection of standard 
model extensions to which it is sensitive, including: R-parity violating interactions in supersymmetric theories, theories with light (electroweak scale) leptoquark degrees of freedom,  non-minimal models of extra doublet Higgs bosons, models in which the quarks and leptons are composite both with and without supersymmetry, and models with strong TeV scale gravitational interactions. Comparing with existing measurements of 
$R_{\pi}$ we provide limits on each of these classes of models; our calculations also represent state of the art theoretical benchmarks 
against which the results from the upcoming round of leptonic pion 
decay experiments may be compared.}

%\archive{hep-ph/yymmxxx}
%\preprintone{XXX}
%\preprinttwo{XXX}
%\submit{}

%%%%%%%%%%%%%%%%%%%%%%%%%%%%%%%%%%%%%%%%%%%%%%%%%%%%%%%%%%%%%%%%%%%%%%%%
%%%%%%%%%%%%%%%%%%%%%%%%%%%%%%%%%%%%%%%%%%%%%%%%%%%%%%%%%%%%%%%%%%%%%%%%

\newpage

%%%%%%%%%%%%%%%%%%%%%%%%%%%%%%%%%%%%%%%%%%%%%%%%%%%%%%%%%%%%%%%%%%%%%%%%%
%%%%%%%%%%%%%%%%%%%%%%%%%%%%%%%%%%%%%%%%%%%%%%%%%%%%%%%%%%%%%%%%%%%%%%%%%

\section{Introduction}

In the Standard Model, the decay of the charged pion is chirally suppressed due to the $V - A$ structure of the weak interaction.  Measurement of the ratio of branching ratios $\frac{\Gamma(\pi\rightarrow e\nu)}{\Gamma(\pi\rightarrow \mu\nu)}$ thus serves as a test of new physics. In particular, various models predict additional pseudoscalar interactions, to which this ratio is particularly sensitive. These include R-parity violating supersymmetry, leptoquarks, extra Higgs theories, certain (super)compositeness models, and extra dimension models with strong gravity at the TeV scale. Given the good agreement between the Standard Model prediction for $\frac{\Gamma(\pi\rightarrow e\nu)}{\Gamma(\pi\rightarrow \mu\nu)}$ and experiment, limits can be placed on the strengths of these interactions \cite{Haber:1978jt,Donoghue:1978cj,McWilliams:1980kj,Shrock:1980ct,
Shanker:1982nd,Mursula:1982em,Hall:1983id,Buchmuller:1986iq,
Kizukuri:1987wr, Buchmuller:1986zs,Campbell:1986xd,Barger:1989rk,Leurer:1993em,
Davidson:1993qk,Herczeg:1995kd,Dreiner:1997uz,Bhattacharyya:1997vv, Allanach:1999ic,Dreiner:2001kc,Dreiner:2002xg,Herz:2002gq,
Chemtob:2004xr,Barbier:2004ez,Erler:2004cx,Dreiner:2006gu,
RamseyMusolf:2007yb, RamseyMusolf:2006vr,Masiero:2008cb}. 

However, experiments at TRIUMF \cite{TRIUMF} and PSI \cite{PSI} should soon reduce the uncertainty on the experimental $R_\pi$, which is currently 40 times greater than its theoretical counterpart, by about one order of magnitude. In view of these ongoing experiments it is timely to analyze the limits that the leptonic pion decay data already impose on theories of physics beyond the electroweak standard model. A companion purpose of this paper is to work out the formalism necessary to analyze generally the predictions for leptonic pion decay in large classes of models, to provide a reference set of calculations to which data from the ongoing round of experiments may be compared.   

We calculate, with new QCD corrections, the effect of an additional induced pseudoscalar interaction on the pion branching ratio. We then use the result to constrain different models of new physics, and compare our findings with the existing literature. Finally, we discuss our conclusions in light of the prospect of new pion decay experiments.

%%%%%%%%%%%%%%%%%%%%%%%%%%%%%%%%%%%%%%%%%%%%%%%%%%%%%%%%%%%%%%%%%%%%%%
%%%%%%%%%%%%%%%%%%%%%%%%%%%%%%%%%%%%%%%%%%%%%%%%%%%%%%%%%%%%%%%%%%%%%%

\section{Pseudoscalar Interactions And Pion Decay}

Let us consider the effective Lagrangian and matrix element for the process $\pi^\pm \rightarrow l^\pm \nu_l$
in the presence of pseudoscalar interactions (we follow the notation of \cite{Campbell:2003ir} throughout this section). We can set limits on the strength of the pseudoscalar interactions from their interference with tree
level W exchange. Since the pion is a pseudoscalar, we will use the following relations for current matrix elements,
\bea
\label{onshellst}
&&\left<0\left|\bar u \gamma_\mu \gamma_5 d\right| \pi(p) \right> = i \sqrt{2}
f_\pi p_\mu \nonumber \\
&& \left<0\left|\bar u \gamma_5 d\right| \pi(p) \right> = i \sqrt{2} \tilde
f_\pi = i \sqrt{2} \hspace{1mm} \frac{f_\pi m^2_\pi}{m_u + m_d} \nonumber \\
%&& \left<0\left|\bar u \sigma^{\mu \nu}\gamma_5 d\right| \pi(p) \right> =0
%\nonumber \\
%&& \left<0\left|\bar u \sigma^{\mu \nu} d\right| \pi(p) \right> =0,
\eea
where $f_\pi = 93 \hspace{1mm}\textnormal{MeV}$ and
$\tilde f_\pi \simeq 1.8 \times 10^5 \hspace{1mm}\textnormal{MeV}^2$.
The matrix element for the tree level W contribution can easily be constructed by using eq.(\ref{onshellst}),
giving;
\be
\mathcal{M}_{W^\pm} = G_F f_\pi \cos \theta_c [\bar l \gamma^\mu(1-\gamma_5)
\nu_l] p_\mu,
\ee
where $p_\mu$ is the pion momentum and $\theta_c$ is the Cabibbo angle.
A pseudoscalar contribution with left-handed neutrinos in the final state
can be expressed as a four-fermi contact operator,
\be
\mathcal{L}_P = -i \frac{\rho}{2 \Lambda^2}[\bar l (1-\gamma_5)\nu_l][\bar u \gamma_5 d]
\ee
where $\rho$ is the pseudoscalar coupling constant. This expression can be
converted to a matrix element using eq.(\ref{onshellst}),
\be
\mathcal{M}_P = \rho \frac{\tilde f_\pi}{\sqrt{2} \Lambda^2}[\bar l (1-\gamma_5) \nu_l].
\ee
In the presence of a pseudoscalar interaction, the overall matrix element for the process $\pi^\pm \rightarrow l^\pm \nu_l$ is the coherent sum,
$\mathcal{M}_{P}+ \mathcal{M}_{W^\pm}=\mathcal{M}_l$.
\be
\mathcal{M}_l= G_F f_\pi \cos \theta_c[\bar l \gamma^\mu(1-\gamma_5)\nu_l]p_\mu + \frac{\rho \tilde
f_\pi}{\sqrt{2} \Lambda^2}[\bar l (1-\gamma_5)\nu_l]
\ee
Having constructed the matrix element, we can now estimate the ratio of branching
ratios,
\be
R_{\pi} \equiv \frac{\Gamma(\pi^- \rightarrow e \nu_e)}{\Gamma(\pi^- \rightarrow \mu \nu_\mu)} =
\frac{(m_\pi^2 -m_e^2)}{(m_\pi^2-m_\mu^2)} \frac{\left<|M_{e
\nu}|^2\right>}{\left<|M_{\mu\nu}|^2\right>}.
\ee
Summing over final states of the squared matrix element we have
\bea
\left<|\mathcal{M}_l|^2\right> &=& 4 \hspace{.5mm} G_f^2 f_\pi^2 \cos^2 \theta_c m_l^2(m_\pi^2-m_l^2)
+ 8 \frac{ G_F \tilde f_\pi f_\pi \cos \theta_c \rho}{\sqrt{2} \Lambda^2}m_l(m_\pi^2 -m_l^2) \nonumber \\
&&+ 2\frac{\rho^2 \tilde f_\pi^2}{\Lambda^4} (m_\pi^2-m_l^2).
\eea
For simplicity we have assumed that the pseudoscalar coupling is
real, however, in general $\rho$ may be complex. The more general expression is obtained by making the following substitution,
\bea
\rho &\rightarrow& \frac{\rho+\rho^*}{2} = \textnormal{Re}(\rho) \nonumber \\
(\rho)^2 &\rightarrow& |\rho|^2.
\eea
We find that the branching ratio is given by
\be
\label{br}
R_{\pi}
=\frac{(m_\pi^2 -m_e^2)}{(m_\pi^2-m_\mu^2)} \left[\frac{m_e^2(m_\pi^2-m_e^2)
+R_e}{m_\mu^2(m_\pi^2-m_\mu^2) + R_\mu} \right],
\ee
where the $R_{e,\mu}$ functions are
\be
\label{Semu}
R_{e,\mu} = \sqrt{2} \frac{\tilde f_\pi \mathrm{Re}(\rho)}{G_F f_\pi \Lambda^2 \cos \theta_c} m_{e,\mu}(m_\pi^2
-m_{e,\mu}^2) + \frac{|\rho|^2 \tilde f_\pi^2}{2 f_\pi^2 G_F^2 \Lambda^4 \cos^2 \theta_c}(m_\pi^2-m_{e,\mu}^2).
\ee

Thus far we have only discussed interactions with left-handed
neutrinos in the final state. The inclusion of right-handed
neutrinos, or neutrinos of a different flavour, requires a modification since pseudoscalar contributions to decays with right-handed, or distinctly flavoured, neutrinos in the final state cannot
interfere with the W exchange graph; hence the contributions to
the rate add incoherently. With right-handed, or distinctly flavoured neutrinos, the expression for the matrix element becomes,
\be \mathcal{M}_P=
\frac{\rho^\prime \tilde f_\pi}{\sqrt{2} \Lambda^2} [\bar l
(1+\gamma_5)\nu_l], \ee where $\rho^\prime$ is the pseudoscalar
coupling involving right-handed neutrinos. Defining 
\be  
R_{\pi}(SM) \equiv
\frac{(m_\pi^2 -m_e^2)^2}{(m_\pi^2-m_\mu^2)^2}
\frac{m_e^2}{m_\mu^2} = 1.28 \space \times \space 10^{-4},
\ee
we can express the branching ratio as
\bea
\label{BRemu} R_{\pi} = R_{\pi}(SM) 
\left(\frac{1 + \sqrt{2}\frac{\tilde f_\pi
\mathrm{Re}(\rho_e)}{G_F \Lambda^2 f_\pi \cos \theta_c m_e} +
\frac{|\rho_e|^2 \tilde f_\pi^2} {2 G_F^2 \Lambda^4 f_\pi^2 \cos^2
\theta_c m_{e}^2} + \frac{|\rho_e^\prime|^{2} \tilde
f_\pi^2}{2 G_F^2 f_\pi^2 \Lambda^4 \cos^2 \theta_c m_e^2}}{1 +
\sqrt{2}\frac{\tilde f_\pi \mathrm{Re}(\rho_\mu)}{G_F \Lambda^2
f_\pi \cos \theta_c m_\mu} + \frac{|\rho_\mu|^2 \tilde f_\pi^2} {2
G_F^2 \Lambda^4 f_\pi^2 \cos^2 \theta_c m_{\mu}^2} +
\frac{|\rho_\mu^\prime|^2 \tilde f_\pi^2}{2 G_F^2 \Lambda^4
f_\pi^2\cos^2 \theta_c m_\mu^2}} \right)
\eea

If we assume either universal scalar couplings or else scalar couplings involving only
the first generation, we would obtain the following approximation for
the ratio of decay widths,
%%%%%%%%%%%%%%%%%%%%%%%%%%%%%%%%%%%%%%%%%%%%%%%%%%%%%%%%%%%%%%%%%%%
\bea
\label{BRa} R_{\pi} \approx R_{\pi}(SM) 
\left(1 + \sqrt{2}\frac{\tilde f_\pi
\mathrm{Re}(\rho)}{G_F \Lambda^2 f_\pi \cos \theta_c m_e} +
\frac{|\rho|^2 \tilde f_\pi^2} {2 G_F^2 \Lambda^4 f_\pi^2 \cos^2
\theta_c m_{e}^2} + \frac{|\rho^\prime|^{2} \tilde f_\pi^2}{2
G_F^2 \Lambda^4 f_\pi^2 \cos^2 \theta_c m_e^2} \right) \nonumber \\
&&
\eea
%%%%%%%%%%%%%%%%%%%%%%%%%%%%%%%%%%%%%%%%%%%%%%%%%%%%%%%%%%%%%%%%%%%

If we only keep the lowest order coherent interference terms in the
couplings $\rho_{e}$ and  $\rho_{\mu}$ the leptonic pion decay ratio would
change to approximately:
\begin{equation}
R_\pi \approx R_\pi\mathrm{(SM)} \left(1 + \sqrt{2}\frac{\tilde f_\pi
\mathrm{Re}(\rho_e)}{G_F \Lambda^2 f_\pi m_e} - \sqrt{2}\frac{\tilde f_\pi
\mathrm{Re}(\rho_\mu)}{G_F \Lambda^2 f_\pi m_\mu}\right),
\end{equation}
The best current theoretical estimate for this ratio in the Standard Model is $R_\pi\mathrm{(SM)} = (1.2352 \pm 0.0001) \cdot 10^{-4}$, and was obtained by calculating QED radiative corrections to the decay \cite{Marciano:1993sh, Decker:1994ea}, and considering higher order terms in the QCD chiral Lagrangian \cite{Cirigliano:2007ga}. Conversely, experiments measure $R_\pi = (1.230 \pm 0.004) \cdot 10^{-4}$ \cite{Yao:2006px,Britton:1992pg,Britton:1993cj, Czapek:1993kc}. 
Combining the errors, we find that at $2\sigma$,
\begin{equation}
\label{eqn:limnoqcd}
-1.07 \cdot 10^{-2} < \sqrt{2}\frac{\tilde f_\pi \mathrm{Re}(\rho_e)}{G_F \Lambda^2 f_\pi m_e} - \sqrt{2}\frac{\tilde f_\pi \mathrm{Re}(\rho_\mu)}{G_F \Lambda^2 f_\pi m_\mu} < 2.27 \cdot 10^{-3}.
\end{equation}

%%%%%%%%%%%%%%%%%%%%%%%%%%%%%%%%%%%%%%%%%%%%%%%%%%%%%%%%%%%%%%%%%%%%%
The additional interactions that we have considered so far interfere coherently with the Standard Model amplitude for pion decay, and the strength of the limit in equation (\ref{eqn:limnoqcd}) is partly due to this interference. However, we may also bound interactions that add incoherently with pion decay. For instance, the terms
\begin{equation}
 -i\frac{\rho'_e}{2\Lambda^2}[\bar{e}(1-\gamma_5)\nu_x][\bar{u}\gamma_5d] - i\frac{\rho'_\mu}{2\Lambda^2}[\bar{\mu}(1-\gamma_5)\nu_y][\bar{u}\gamma_5d],
\end{equation}
where $x \neq e$ and $y \neq \mu$, contribute to pion decay but only add incoherently with the Standard Model amplitude. However, the lack of chiral suppression still allows us to put competitive bounds on some of these terms. These interactions would change the pion branching ratio to
%~\cite{Campbell:2003ir}
\begin{equation}
 R_\pi \approx R_\pi\mathrm{(SM)} \left(1 + \frac{\tilde f_\pi^2
|\rho'_e|^2}{2 G_F^2 \Lambda^4 f_\pi^2 m_e^2} - \frac{\tilde f_\pi^2
|\rho'_\mu|^2}{2 G_F^2 \Lambda^4 f_\pi^2 m_\mu^2}\right).
\end{equation}
%%%%%%%%%%%%%%%%%%%%%%%%%%%%%%%%%%%%%%%%%%%%%%%%%%%%%%%%%%%%%%%%%%%%%%%
Following the above analysis we obtain the limit
\begin{equation} \label{limit2}
-1.07 \cdot 10^{-2} < \frac{\tilde f_\pi^2 |\rho'_e|^2}{2 G_F^2 \Lambda^4 f_\pi^2 m_e^2} - \frac{\tilde f_\pi^2
|\rho'_\mu|^2}{2 G_F^2 \Lambda^4 f_\pi^2 m_\mu^2} < 2.27 \cdot 10^{-3}.
\end{equation}
%%%%%%%%%%%%%%%%%%%%%%%%%%%%%%%%%%%%%%%%%%%%%%%%%%%%%%%%%%%%%%%%%%%%%%

The above limits were obtained ignoring QCD renormalization.
We now calculate the one-loop QCD correction to these limits, still treating the additional interaction as a four-fermion local operator. The anomalous dimension of the operator $[\bar{e}(1-\gamma_5)\nu_e][\bar{u}\gamma_5d]$ is $\gamma = -\frac{2}{\pi}\alpha_s$ 
We may use this anomalous dimension to compute the QCD enhancement of a pseudoscalar contribution to pion decay, regardless of the underlying physics. First, recall that we may write the running of an operator coefficient as
\begin{equation}
 c(p_1) = \left( \frac{\alpha(p_1)}{\alpha(p_2)} \right)^\frac{a_\mathcal{O}}{2 b_0} c(p_2).
\end{equation}
For the pseudoscalar pion decay operator, $a_\mathcal{O} = 8$. Now, assume that the new physics responsible for this operator occurs at a scale $\Lambda \sim 1~\mathrm{TeV}$. From $\Lambda$ down to $m_t$, we have six quarks, so
\begin{equation}
 b_0 = 11 - \frac{2}{3} (6) = 7,
\end{equation}
from the QCD beta function. QCD running from $\Lambda$ to $m_t$ thus enhances the operator by a factor of
\begin{equation}
 \frac{c(m_t)}{c(\Lambda)} = \left( \frac{\alpha_s(m_t)}{\alpha_s(\Lambda)} \right)^\frac{4}{7}.
\end{equation}
Note that since $\alpha_s$ \emph{decreases} with increasing energy, this is indeed an enhancement. From $m_t$ down to $m_b$, the renormalization group running again enhances the pseudoscalar operator. However, since the top quark is integrated out below $m_t$, we must now use
\begin{equation}
 b_0 = 11 - \frac{2}{3} (5) = \frac{23}{3}.
\end{equation}
This procedure may be continued all the way down to the scale $4 \pi f_\pi$, where QCD becomes strongly coupled and chiral symmetry is spontaneously broken \cite{Weinberg:1978kz,Georgi:1985kw}. At this point, perturbation theory no longer provides a good description of the physics, and thus cannot predict the renormalization group running of the pseudoscalar pion decay operator.

By multiplying the enhancement factors from each stage of the running, however, we can at least obtain a partial estimate of the effects of renormalization on our limits. We find that the leading QCD corrections will enhance each bound by a factor of 
\begin{equation}
 \label{eqn:qcdcorr}
 \left(\frac{\alpha_s(4\pi f_\pi)}{\alpha_s(m_c)}\right)^{4/9} \left(\frac{\alpha_s(m_c)}{\alpha_s(m_b)}\right)^{12/25} \left(\frac{\alpha_s(m_b)}{\alpha_s(m_t)}\right)^{12/23} \left(\frac{\alpha_s(m_t)}{\alpha_s(\Lambda)}\right)^{4/7},
\end{equation}
where $\Lambda$ is the scale at which the interaction occurs. Each factor may be computed numerically using the renormalization group equation for $\alpha_s$, given the value of $\alpha_s$ at some reference point. We take $\alpha_s(m_Z) = 0.1176$, $m_c = 1.27~\mathrm{GeV}$, $m_b = 4.20~\mathrm{GeV}$, and $m_t = 171.2~\mathrm{GeV}$ \cite{Yao:2006px}, and run $\alpha_s$ using the QCD beta function. We also choose $\Lambda = 1~\mathrm{TeV}$, to obtain a conservative lower bound on the QCD enhancement of any new operator. (Although the enhancement would be greater if $\Lambda$ were increased, the difference would be minimal, since $\alpha_s$ runs very slowly at high energies.) Since the result is approximately $2.0$, the net effect of the renormalization group flow has been to double the significance of any pseudoscalar contribution to pion decay. Conversely, given an experimental limit on the size of the pseudoscalar operator contribution to pion decay, our computation of the QCD running will allow us to place limits on possible sources of pseudoscalar contributions to pion decay that are twice as strong. The new limit is thus given by 
\begin{equation} \label{limit}
-5.3 \cdot 10^{-3} < \sqrt{2}\frac{\tilde f_\pi \mathrm{Re}(\rho_e)}{G_F \Lambda^2 f_\pi m_e} - \sqrt{2}\frac{\tilde f_\pi \mathrm{Re}(\rho_\mu)}{G_F \Lambda^2 f_\pi m_\mu} < 1.1 \cdot 10^{-3}.
\end{equation}
Similarly, for the incoherent contributions to pion decay, after including QCD corrections we obtain the limits:
\begin{equation}
-5.3 \cdot 10^{-3} < \frac{\tilde f_\pi^2 |\rho'_e|^2}{2 G_F^2 \Lambda^4 f_\pi^2 m_e^2} - \frac{\tilde f_\pi^2
|\rho'_\mu|^2}{2 G_F^2 \Lambda^4 f_\pi^2 m_\mu^2} < 1.1 \cdot 10^{-3}.
\end{equation}

%%%%%%%%%%%%%%%%%%%%%%%%%%%%%%%%%%%%%%%%%%%%%%%%%%%%%%%%%%%%%%%%%%%%%%%
%%%%%%%%%%%%%%%%%%%%%%%%%%%%%%%%%%%%%%%%%%%%%%%%%%%%%%%%%%%%%%%%%%%%%%%

\section{Supersymmetry}

%%%%%%%%%%%%%%%%%%%%%%%%%%%%%%%%%%%%%%%%%%%%%%%%%%%%%%%%%%%%%%%%%%%%%%%

\subsection{R-Parity Conserving Supersymmetry}

%%%%%%%%%%%%%%%%%%%%%%%%%%%%%%%%%%%%%%%%%%%%%%%%%%%%%%%%%%%%%%%%%%%%%%%%%%
%Corrections to the Higgs propagator from the particles in the Standard 
%Model cause divergent shifts in the Higgs mass, which should be cut off 
%at the scale of new physics. If there is no new physics until the Planck 
%scale, then an extremely precise cancellation would have to take place
% between the Higgs mass corrections to give the mass that is implied by 
%experimental data. Now, fermions and bosons lead to Higgs mass 
%corrections of opposite sign. Supersymmetry exploits this sign difference 
%to resolve the hierarchy problem by proposing a symmetry between fermions 
%and bosons. 
%%%%%%%%%%%%%%%%%%%%%%%%%%%%%%%%%%%%%%%%%%%%%%%%%%%%%%%%%%%%%%%%%%%%%%%%
In supersymmetric models, for each Standard Model fermion, there is a boson, and vice versa. These superpartners have the same quantum numbers as the original particles, except for spin.
%%%%%%%%%%%%%%%%%%%%%%%%%%%%%%%%%%%%%%%%%%%%%%%%%%%%%%%%%%%%%%%%%%%%%%%%
Both gauge and nongauge interactions in globally supersymmetric theories may be most easily described in terms of interaction terms expressed in terms of superfields, each containing a Standard Model field with its superpartner (along with an auxiliary field) \cite{Ramond:1999vh,Wess:1992cp}. Using this formalism, we may construct the superpotential which encodes all the nongauge interactions of the minimal supersymmetric extension of the standard model.
\begin{equation}
 W = u_{ij} (Q_L^i H_u) U_L^{j,c} + d_{ij} (Q_L^i H_d) D_L^{j,c} + e_{ij} (L_L^i H_d) E_L^{j,c} + \mu H_u H_d,
\end{equation}
where the superfield $Q_L^i$ contains the left-handed $i$th-generation quark $SU(2)$ doublet, $E_L^{j,c}$ contains the conjugate of the right-handed $j$th-generation charged lepton $SU(2)$ singlet, etc. Superpotentials are always holomorphic functions of the superfields; i.e., they never contain the complex conjugates of superfields. We have used parentheses to indicate implicit contraction of $SU(2)$ indices, and omitted $SU(3)$ indices. Note that supersymmetric models have two doublets of Higgs fields; here, they are denoted by $H_u$ and $H_d$, for the quarks to which they couple. $W$ is the superpotential for the Minimal Supersymmetric Standard Model (MSSM), the simplest extension of the Standard Model incorporating supersymmetry. 

However, the MSSM superpotential does not contain all of the possible terms allowed by gauge invariance \cite{Ramond:1999vh,Martin:1997ns}.
By itself gauge invariance would allow dimension four and higher terms which violate either or both of baryon number or lepton number. (Note that these are global symmetries, not gauge symmetries.) 
To preserve baryon and lepton number, which prevent experimentally constrained processes such as proton decay from occurring, one conventionally imposes a discrete $Z_2$ symmetry called R-parity, under which a particle has charge
\begin{equation}
 R = (-1)^{3B + L + 2S},
\end{equation}
where $B$, $L$, and $S$ are baryon number, lepton number, and spin, respectively. With this definition, all Standard Model particles have $R = 1$, and their superpartners have $R = -1$.
With the assumption of MSSM field content, gauge invariance, and R-parity conservation, the superpotential governing nongauge interactions in the MSSM only contains the terms displayed above.

%%%%%%%%%%%%%%%%%%%%%%%%%%%%%%%%%%%%%%%%%%%%%%%%%%%%%%%%%%%%%%%%%%%%%%
R-parity conserving supersymmetry, as implemented in the Minimal Supersymmetric Standard Model (MSSM), is unfortunately only weakly constrained by the precision measurement of the ratio of leptonic branching ratios $R_{\pi}$ in pion decay. Although the model contains two Higgs doublets, and hence charged Higgs scalars, the fact that only one of the doublets couples to leptons, providing the charged lepton mass, means that the coupling of the physical charged Higgs to leptons is lepton mass proportional, and hence the effect of their tree-level exchange will cancel in the ratio $R_{\pi}$. MSSM supersymmetric loop induced contributions, to either an induced pseudoscalar, or axial-vector, amplitude in leptonic pion decay have been computed in \cite{RamseyMusolf:2007yb}. They are not at a level that would have been detected in present experiments; furthermore, in order that they be sufficiently large to be detectable in the upcoming round of pion decay experiments, one would need to choose the supersymmetry breaking parameters to be sufficiently small that the superpartners would be copiously produced at the LHC, and generally that their direct discovery would be possible with initial data at the LHC. 
While the R-parity conserving MSSM is more accessible to direct discovery at colliders, rather than through its effects in  $R_{\pi}$, when one allows for the possibility of R-parity violation then pion decay again becomes a potentially powerful probe, as we now discuss. 
%%%%%%%%%%%%%%%%%%%%%%%%%%%%%%%%%%%%%%%%%%%%%%%%%%%%%%%%%%%%%%%%%%%%%%

\subsection{R-Parity Violating Supersymmetry}

\label{sect:rpvsusy}

%%%%%%%%%%%%%%%%%%%%%%%%%%%%%%%%%%%%%%%%%%%%%%%%%%%%%%%%%%%%%%%%%%%%%%%%%
As noted in the previous subsection, the MSSM superpotential does not contain all of the possible terms allowed by gauge invariance \cite{Ramond:1999vh,Martin:1997ns}. In particular, the $L_L$ and $H_d$ superfields have the same gauge quantum numbers, and we may thus make the substitution $H_d \rightarrow L_L$ in any allowed superpotential term to obtain another allowed superpotential term. This substitution may be performed on the second and third terms in the MSSM superpotential above. With the last term, the substitution gives an unphysical, after field redefinition, bilinear coupling between leptons and Higgs fields, which we omit. Also, a $U_L^c D_L^c D_L^c$ term is not forbidden by any of the gauge symmetries. The additional superpotential terms that we consider are thus of the form
\begin{equation}
 \lambda_{ijk} (L_L^i L_L^j) E_L^{k,c} + \lambda'_{ijk} (L_L^i Q_L^j) D_L^{k,c} + \lambda''_{ijk} U_L^{i,c} D_L^{j,c} D_L^{k,c},
\end{equation}
where again $SU(2)$ singlets are indicated by parentheses and $SU(3)$ indices are omitted. Note that by the $SU(2)$ and $SU(3)$ gauge symmetries, respectively, we have $\lambda_{ijk} = -\lambda_{jik}$ and $\lambda''_{ijk} = -\lambda''_{ikj}$. 
Each of these R-parity violating terms explicitly violates either baryon number or lepton number. 
%%%%%%%%%%%%%%%%%%%%%%%%%%%%%%%%%%%%%%%%%%%%%%%%%%%%%%%%%%%%%%%%%%%%%%%

\begin{figure}
\begin{center}
\includegraphics[width=\textwidth]{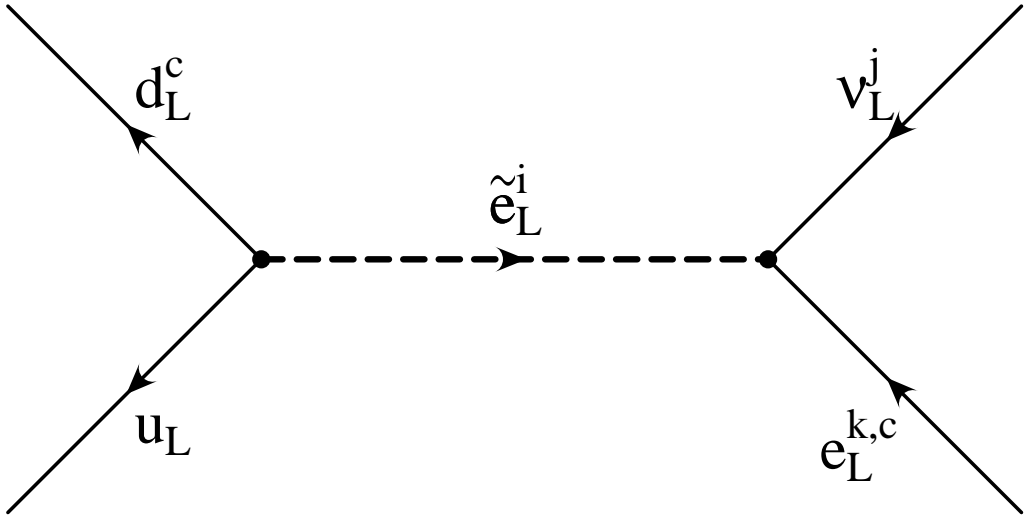}
\caption[RPV SUSY slepton exchange contribution to pion decay]{An R-parity violating SUSY contribution to pion decay. The superpartner of the left-handed $i$th generation charged lepton is denoted by $\tilde{e}^i_L$.}
\label{fig:rpvsusy1}
\end{center}
\end{figure}

\begin{figure}
\begin{center}
\includegraphics[height=0.5\textheight]{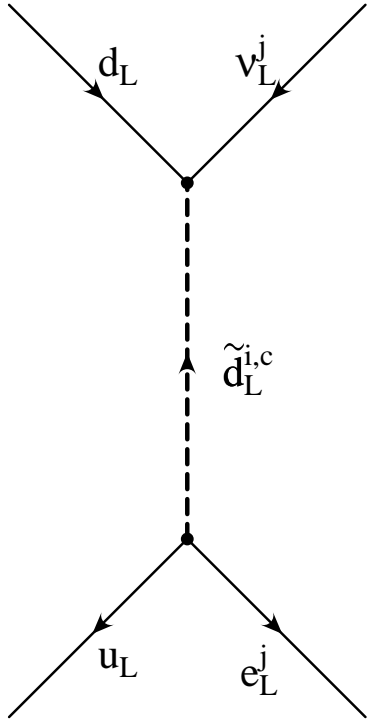}
\caption[RPV SUSY squark exchange contribution to pion decay]{Another R-parity violating SUSY contribution to pion decay, which has been considered before~\cite{Barger:1989rk}. The superpartner of the left-handed $i$th generation down quark is denoted by $\tilde{d}^i_L$.}
\label{fig:rpvsusy2}
\end{center}
\end{figure}

If the $B$-violating $U_L^c D_L^c D_L^c$ and $L$-violating $(L_L Q_L) D_L^c$ terms are both allowed, proton decay places extremely stringent constraints on the products $\lambda' \lambda''$~\cite{Smirnov:1996bg}. Other bounds involving the $B$-violating couplings are discussed in 
\cite{Carlson:1995ji}; in particular, when the $U_L^c D_L^c D_L^c$ and $(L_L L_L) E_L^c$ terms are allowed, proton decay still yields excellent constraints on most products of couplings. We consider only the $L$-violating terms, setting $\lambda''_{ijk} = 0$. Under different coupling structures, two types of contributions to pion decay are possible, shown in figures \ref{fig:rpvsusy1} and \ref{fig:rpvsusy2}. We assume the double coupling dominance hypothesis, considering only two couplings to be significant at a time. This assumption allows to apply our general limits from the previous section to each new contribution separately, ignoring cancellations between the contributions.

Two types of diagrams are allowed by the new superpotential terms. The first, shown in figure \ref{fig:rpvsusy1}, 
corresponds to slepton exchange and has effective strength
\begin{align}
\rho_k &= \frac{\mathrm{Re}(\lambda_{ikk}\lambda_{i11}^{'*})}{2} \nonumber \\
\rho'_k &= \frac{\mathrm{Re}(\lambda_{ijk}\lambda_{i11}^{'*})}{2},
\end{align}
where $\lambda$ and $\lambda'$ are the couplings at the vertices, and in the second equation $j \neq k$. From this, we can obtain a limit on the product of couplings in terms of the mass of the mediating particle. The limits obtained in this manner are shown in table \ref{tab:rpvsusy1}. Note that throughout our results, we use $[\tilde{q}]$ to represent $\frac{m_{\tilde{q}}}{100\ \mathrm{GeV}}$.
\begin{table}
\begin{center}
\begin{tabular}{|c|c|c|}
\hline
Product of couplings & New limit & Previous limit from \cite{Herz:2002gq}\\
\hline
$\mathrm{Re}(\lambda_{121}\lambda_{111}^{'*})$ & $2.9 \times 10^{-6} [\tilde{e}_1]^{2}$ & \\
$\mathrm{Re}(\lambda_{122}\lambda_{111}^{'*})$ & $4.8 \times 10^{-5} [\tilde{e}_1]^{2}$ & ${}^* 2.0 \times 10^{-5} [\tilde{e}_1]^{2}$ \\
$\mathrm{Re}(\lambda_{121}\lambda_{211}^{'*})$ & $2.3 \times 10^{-7} [\tilde{e}_2]^{2}$ & $5.0 \times 10^{-7} [\tilde{e}_2]^{2}$ \\
$\mathrm{Re}(\lambda_{122}\lambda_{211}^{'*})$ & $1.3 \times 10^{-3} [\tilde{e}_2]^{2}$ & \\
$\mathrm{Re}(\lambda_{131}\lambda_{111}^{'*})$ & $2.9 \times 10^{-6} [\tilde{e}_1]^{2}$ & \\
$\mathrm{Re}(\lambda_{132}\lambda_{111}^{'*})$ & $1.3 \times 10^{-3} [\tilde{e}_1]^{2}$ & \\
$\mathrm{Re}(\lambda_{131}\lambda_{311}^{'*})$ & $2.3 \times 10^{-7} [\tilde{e}_3]^{2}$ & $5.0 \times 10^{-7} [\tilde{e}_3]^{2}$ \\
$\mathrm{Re}(\lambda_{132}\lambda_{311}^{'*})$ & $1.3 \times 10^{-3} [\tilde{e}_3]^{2}$ & \\
$\mathrm{Re}(\lambda_{231}\lambda_{211}^{'*})$ & $2.9 \times 10^{-6} [\tilde{e}_2]^{2}$ & \\
$\mathrm{Re}(\lambda_{232}\lambda_{211}^{'*})$ & $1.3 \times 10^{-3} [\tilde{e}_2]^{2}$ & \\
$\mathrm{Re}(\lambda_{231}\lambda_{311}^{'*})$ & $2.9 \times 10^{-6} [\tilde{e}_3]^{2}$ & \\
$\mathrm{Re}(\lambda_{232}\lambda_{311}^{'*})$ & $4.8 \times 10^{-5} [\tilde{e}_3]^{2}$ & ${}^* 2.0 \times 10^{-5} [\tilde{e}_3]^{2}$ \\
\hline
\end{tabular}
\caption[Limits on RPV SUSY couplings from slepton exchange contribution]{Limits on R-parity violating SUSY coupling combinations $\mathrm{Re}(\lambda\lambda^{'*})$ from slepton exchange. The starred entries are previous limits from reference \cite{Herz:2002gq} which we believe to be erroneous, as they seem to have been obtained under the assumption that the interference from amplitudes that add coherently is always constructive. We use $[\tilde{e}]$ to represent $\frac{m_{\tilde{e}}}{100\ \mathrm{GeV}}$.}
\label{tab:rpvsusy1}
\end{center}
\end{table}

We may obtain another effective contribution to pion decay by Fierz reordering an interaction arising purely from the $L_L Q_L D_L^c$ couplings, shown in figure \ref{fig:rpvsusy2}. This contribution has already been considered \cite{Barger:1989rk,Herz:2002gq,RamseyMusolf:2007yb}, but we update the limits using current experimental data and the QCD correction. The Fierz reordering weakens the bound, since there is now chiral suppression. We have
\begin{equation}
R_\pi \approx R_\pi\mathrm{(SM)} \left(1 + \left( \frac{1}{2 \sqrt{2} G_F m^2} \right) \left( |\lambda'_{11i}|^2 - |\lambda'_{21i}|^2 \right) \right),
\end{equation}
where $m$ is the mass of the mediating down squark. The resulting limits are shown in table \ref{tab:rpvsusy2}. 

\begin{table}
\begin{center}
\begin{tabular}{|c|c|c|}
\hline
Product of couplings & New limit & Previous limit from \cite{Herz:2002gq} \\
\hline
$|\lambda'_{111}|^2$ & $3.6 \times 10^{-4} [\tilde{d}_1]^{2}$ & $6.8 \times 10^{-4} [\tilde{d}_1]^{2}$ \\
$|\lambda'_{112}|^2$ & $3.6 \times 10^{-4} [\tilde{d}_2]^{2}$ & $6.8 \times 10^{-4} [\tilde{d}_2]^{2}$ \\
$|\lambda'_{113}|^2$ & $3.6 \times 10^{-4} [\tilde{d}_3]^{2}$ & $6.8 \times 10^{-4} [\tilde{d}_3]^{2}$ \\
$|\lambda'_{211}|^2$ & $1.7 \times 10^{-3} [\tilde{d}_1]^{2}$ & $3.5 \times 10^{-3} [\tilde{d}_1]^{2}$ \\
$|\lambda'_{212}|^2$ & $1.7 \times 10^{-3} [\tilde{d}_2]^{2}$ & $3.5 \times 10^{-3} [\tilde{d}_2]^{2}$ \\
$|\lambda'_{213}|^2$ & $1.7 \times 10^{-3} [\tilde{d}_3]^{2}$ & $3.5 \times 10^{-3} [\tilde{d}_3]^{2}$ \\
\hline
\end{tabular}
\caption[Limits on RPV SUSY couplings from squark exchange contribution]{Updated limits on R-parity violating SUSY couplings $|\lambda'|^2$ from squark exchange. We use $[\tilde{q}]$ to represent $\frac{m_{\tilde{q}}}{100\ \mathrm{GeV}}$.}
\label{tab:rpvsusy2}
\end{center}
\end{table}

%%%%%%%%%%%%%%%%%%%%%%%%%%%%%%%%%%%%%%%%%%%%%%%%%%%%%%%%%%%%%%%%%%%%%%%
%%%%%%%%%%%%%%%%%%%%%%%%%%%%%%%%%%%%%%%%%%%%%%%%%%%%%%%%%%%%%%%%%%%%%%%

\section{Leptoquarks}
\label{sect:lq}
Leptoquarks, particles that have interactions vertices that simultaneously involve both leptons and quarks, appear in various extensions of the Standard Model, including Grand Unified Theories 
\cite{Ross:1985ai,Pati:1974yy,Georgi:1974sy} 
and technicolour models \cite{Farhi:1980xs}. If we require that the Lagrangian remains invariant under $SU(3) \times SU(2) \times U(1)$ transformations, but allow violation of the global symmetries corresponding to baryon and lepton number, we may write the new terms \cite{Buchmuller:1986zs}
\begin{align}
 \label{eqn:lq}
 \mathcal{L} &\supset (\lambda^{ij}_{LS_0} \bar{Q}_L^{j,c} i \sigma^2 L_L^i + \lambda^{ij}_{RS_0} \bar{u}_R^{j,c} e_R^i) S_0^\dag + \lambda^{ij}_{R\tilde{S}_0} \bar{d}_R^{j,c} e_R^i \tilde{S}_0^\dag \nonumber \\
 &+ (\lambda^{ij}_{LS_{1/2}} \bar{u}_R^j L_L^i + \lambda^{ij}_{RS_{1/2}} \bar{Q}_L^j i \sigma^2 e_R^i) S_{1/2}^\dag + \lambda^{ij}_{L\tilde{S}_{1/2}} \bar{d}_R^j L_L^i \tilde{S}_{1/2}^\dag + \lambda^{ij}_{LS_1} \bar{Q}_L^{j,c} i \sigma^2 \sigma^a L_L^i S_1^{a\dag} \nonumber \\
 &+ (\lambda^{ij}_{LV_0} \bar{Q}_L^j \gamma_\mu L_L^i + \lambda^{ij}_{RV_0} \bar{d}_R^j \gamma_\mu e_R^i) V_0^{\mu\dag} + \lambda^{ij}_{R\tilde{V}_0} \bar{u}_R^j \gamma_\mu e_R^i \tilde{V}_0^{\mu\dag} \nonumber \\
 &+ (\lambda^{ij}_{LV_{1/2}} \bar{d}_R^{j,c} \gamma_\mu L_L^i + \lambda^{ij}_{RV_{1/2}} \bar{Q}_L^{j,c} \gamma_\mu e_R^i) V_{1/2}^{\mu\dag} + \lambda^{ij}_{L\tilde{V}_{1/2}} \bar{u}_R^{j,c} \gamma_\mu L_L^i \tilde{V}_{1/2}^{\mu\dag} \nonumber \\
 &+ \lambda^{ij}_{LV_1} \bar{Q}_L^j \gamma_\mu \sigma^a L_L^i V_1^{a\mu\dag} +~\mathrm{h.c.},
\end{align}
where $X^c$ is the conjugate of the field $X$, and barred fields are outgoing. (Note that in this convention, which we have adopted from \cite{Buchmuller:1986zs}, $u_R^c$ is the conjugate of the $u_R$ field, which is \emph{left-handed}. This differs from the convention used everywhere else in this work, where $u_R^c$ is the \emph{right-handed} up anti-quark field.) A summary of the leptoquarks and their quantum numbers is given in table \ref{tab:lq}. We have explicitly indicated the $SU(2)$ algebra of weak isospin using the Pauli matrices.

\begin{table}
\begin{center}
\begin{tabular}{|c|c|c|c|}
\hline
Leptoquark & Spin & Charge & Hypercharge \\
\hline
$S_0$ & $0$ & $-\frac{1}{3}$ & $-\frac{2}{3}$ \\
$\tilde{S}_0$ & $0$ & $-\frac{4}{3}$ & $-\frac{8}{3}$ \\
$S_{1/2}$ & $0$ & $-\frac{2}{3}, -\frac{5}{3}$ & $-\frac{7}{3}$ \\
$\tilde{S}_{1/2}$ & $0$ & $\frac{1}{3}, -\frac{2}{3}$ & $-\frac{1}{3}$ \\
$S_1$ & $0$ & $-\frac{2}{3}, -\frac{1}{3}, -\frac{4}{3}$ & $-\frac{2}{3}$ \\
$V_0$ & $1$ & $-\frac{2}{3}$ & $-\frac{4}{3}$ \\
$\tilde{V}_0$ & $1$ & $-\frac{5}{3}$ & $-\frac{10}{3}$ \\
$V_{1/2}$ & $1$ & $-\frac{1}{3}, -\frac{4}{3}$ & $-\frac{5}{3}$ \\
$\tilde{V}_{1/2}$ & $1$ & $\frac{2}{3}, -\frac{1}{3}$ & $\frac{1}{3}$ \\
$V_1$ & $1$ & $\frac{1}{3}, -\frac{2}{3}, -\frac{5}{3}$ & $-\frac{4}{3}$ \\
\hline
\end{tabular}
\caption[Leptoquark quantum numbers]{Quantum numbers of the leptoquarks in equation \ref{eqn:lq}.}
\label{tab:lq}
\end{center}
\end{table}

Leptoquarks can provide both pseudoscalar and axial-vector contributions to pion decay. Both of these types of contributions interfere with the Standard Model amplitude, but the axial-vector contribution is still chirally suppressed, and will thus lead to less stringent bounds. Here, we revise the previous limits on leptoquark couplings, using the latest experimental data for $R_\pi$ and including the QCD correction factor.

Pseudoscalar leptoquark contributions to pion decay arise from the effective four-fermion interactions
\begin{eqnarray}
\mathcal{L}_{\mathrm{eff}} &\supset& \frac{\lambda^{ij}_{LS_0}\lambda^{kl*}_{RS_0}}{2m_{S_0}^2}(\bar{d}_L^{j,c}u_R^{l,c})(\bar{e}_R^k\nu_L^{i}) \\
&+& \frac{\lambda^{ij}_{LS_{1/2}}\lambda^{kl*}_{RS_{1/2}}}{2m_{S_{1/2}}^2}(\bar{u}_R^{j}d_L^l)(\bar{e}_R^k\nu_L^{i}) \\
&+& \frac{\lambda^{ij}_{LV_0}\lambda^{kl*}_{RV_0}}{m_{V_0}^2}(\bar{u}_L^{j}d_R^l)(\bar{e}_R^{k}\nu_L^{i}) \\
&+& \frac{\lambda^{ij}_{LV_{1/2}}\lambda^{kl*}_{RV_{1/2}}}{m_{V_{1/2}}^2}(\bar{d}_R^{j,c}u_L^{l,c})(\bar{e}_R^{k}\nu_L^{i}) +~\mathrm{h.c.}
\end{eqnarray}
(Note that the interactions have been Fierz reordered.) These interactions contribute to pion decay when any of the products $\lambda^{i1}\lambda^{k1}$ is nonzero, where $k = 1$ for the electron decay channel and $k = 2$ for the muon decay channel, and $i$ is the generational index of the unobservable outgoing neutrino. When $i = k$, the contribution interferes coherently with the W exchange amplitude; otherwise, the interference is incoherent. Taking only the real part, we have
\begin{align}
\rho_k = \left(\frac{1}{4}\right) \mathrm{Re}(\lambda^{k1}_{LS} \lambda^{k1*}_{RS}),&~~ \rho'_k = \left(\frac{1}{4}\right) \mathrm{Re}(\lambda^{i1}_{LS} \lambda^{k1*}_{RS}),\\
\rho_k = \left(\frac{1}{2}\right) \mathrm{Re}(\lambda^{k1}_{LV} \lambda^{k1*}_{RV}),&~~ \rho'_k = \left(\frac{1}{2}\right) \mathrm{Re}(\lambda^{i1}_{LV} \lambda^{k1*}_{RV}),
\end{align}
for the contributions involving scalar and vector leptoquarks, respectively, where for the $\rho'$ we have assumed $i \neq k$. Substituting into equation \ref{limit}, we obtain bounds on products of leptoquark couplings, which are shown in table \ref{tab:pslq}.
\begin{table}
\begin{center}
\begin{tabular}{|c|c|c|}
\hline
Product of couplings & New limit & Previous limit from \cite{Davidson:1993qk} or \cite{Herz:2002gq} \\
\hline
$\mathrm{Re}(\lambda^{11}_{LS_0}\lambda^{11*}_{RS_0})$ & $4.6 \times 10^{-7} [S_0]^2$ & $1 \times 10^{-6} [S_0]^2$ \\
$\mathrm{Re}(\lambda^{11}_{LS_0}\lambda^{21*}_{RS_0})$ & $2.6 \times 10^{-3} [S_0]^2$ & ${}^*2 \times 10^{-4} [S_0]^2$ \\
$\mathrm{Re}(\lambda^{21}_{LS_0}\lambda^{11*}_{RS_0})$ & $5.8 \times 10^{-6} [S_0]^2$ & ${}^*1 \times 10^{-6} [S_0]^2$ \\
$\mathrm{Re}(\lambda^{21}_{LS_0}\lambda^{21*}_{RS_0})$ & $9.5 \times 10^{-5} [S_0]^2$ & $2 \times 10^{-4} [S_0]^2$ \\
$\mathrm{Re}(\lambda^{31}_{LS_0}\lambda^{11*}_{RS_0})$ & $5.8 \times 10^{-6} [S_0]^2$ & ${}^*1 \times 10^{-6} [S_0]^2$ \\
$\mathrm{Re}(\lambda^{31}_{LS_0}\lambda^{21*}_{RS_0})$ & $2.6 \times 10^{-3} [S_0]^2$ & ${}^*2 \times 10^{-4} [S_0]^2$ \\
\hline
$\mathrm{Re}(\lambda^{11}_{LS_{1/2}}\lambda^{11*}_{RS_{1/2}})$ & $4.6 \times 10^{-7} [S_{1/2}]^2$ & $1 \times 10^{-6} [S_{1/2}]^2$ \\
$\mathrm{Re}(\lambda^{11}_{LS_{1/2}}\lambda^{21*}_{RS_{1/2}})$ & $2.6 \times 10^{-3} [S_{1/2}]^2$ & ${}^*2 \times 10^{-4} [S_{1/2}]^2$ \\
$\mathrm{Re}(\lambda^{21}_{LS_{1/2}}\lambda^{11*}_{RS_{1/2}})$ & $5.8 \times 10^{-6} [S_{1/2}]^2$ & ${}^*1 \times 10^{-6} [S_{1/2}]^2$ \\
$\mathrm{Re}(\lambda^{21}_{LS_{1/2}}\lambda^{21*}_{RS_{1/2}})$ & $9.5 \times 10^{-5} [S_{1/2}]^2$ & $2 \times 10^{-4} [S_{1/2}]^2$ \\
$\mathrm{Re}(\lambda^{31}_{LS_{1/2}}\lambda^{11*}_{RS_{1/2}})$ & $5.8 \times 10^{-6} [S_{1/2}]^2$ & ${}^*1 \times 10^{-6} [S_{1/2}]^2$ \\
$\mathrm{Re}(\lambda^{31}_{LS_{1/2}}\lambda^{21*}_{RS_{1/2}})$ & $2.6 \times 10^{-3} [S_{1/2}]^2$ & ${}^*2 \times 10^{-4} [S_{1/2}]^2$ \\
\hline
$\mathrm{Re}(\lambda^{11}_{LV_0}\lambda^{11*}_{RV_0})$ & $2.3 \times 10^{-7} [V_0]^2$ & ${}^\dagger 9.8 \times 10^{-8} [V_0]^2$ \\
$\mathrm{Re}(\lambda^{11}_{LV_0}\lambda^{21*}_{RV_0})$ & $1.3 \times 10^{-3} [V_0]^2$ & ${}^*1 \times 10^{-4} [V_0]^2$ \\
$\mathrm{Re}(\lambda^{21}_{LV_0}\lambda^{11*}_{RV_0})$ & $2.9 \times 10^{-6} [V_0]^2$ & ${}^*5 \times 10^{-7} [V_0]^2$ \\
$\mathrm{Re}(\lambda^{21}_{LV_0}\lambda^{21*}_{RV_0})$ & $4.8 \times 10^{-5} [V_0]^2$ & $1.0 \times 10^{-4} [V_0]^2$ \\
$\mathrm{Re}(\lambda^{31}_{LV_0}\lambda^{11*}_{RV_0})$ & $2.9 \times 10^{-6} [V_0]^2$ & ${}^*5 \times 10^{-7} [V_0]^2$ \\
$\mathrm{Re}(\lambda^{31}_{LV_0}\lambda^{21*}_{RV_0})$ & $1.1 \times 10^{-3} [V_0]^2$ & ${}^*1 \times 10^{-4} [V_0]^2$ \\
\hline
$\mathrm{Re}(\lambda^{11}_{LV_{1/2}}\lambda^{11*}_{RV_{1/2}})$ & $2.3 \times 10^{-7} [V_{1/2}]^2$ & $5.0 \times 10^{-7} [V_{1/2}]^2$ \\
$\mathrm{Re}(\lambda^{11}_{LV_{1/2}}\lambda^{21*}_{RV_{1/2}})$ & $1.3 \times 10^{-3} [V_{1/2}]^2$ & ${}^*1 \times 10^{-4} [V_{1/2}]^2$ \\
$\mathrm{Re}(\lambda^{21}_{LV_{1/2}}\lambda^{11*}_{RV_{1/2}})$ & $2.9 \times 10^{-6} [V_{1/2}]^2$ & ${}^*5 \times 10^{-7} [V_{1/2}]^2$ \\
$\mathrm{Re}(\lambda^{21}_{LV_{1/2}}\lambda^{21*}_{RV_{1/2}})$ & $4.8 \times 10^{-5} [V_{1/2}]^2$ & ${}^\dagger 2.0 \times 10^{-5} [V_{1/2}]^2$ \\
$\mathrm{Re}(\lambda^{31}_{LV_{1/2}}\lambda^{11*}_{RV_{1/2}})$ & $2.9 \times 10^{-6} [V_{1/2}]^2$ & ${}^*5 \times 10^{-7} [V_{1/2}]^2$ \\
$\mathrm{Re}(\lambda^{31}_{LV_{1/2}}\lambda^{21*}_{RV_{1/2}})$ & $1.3 \times 10^{-3} [V_{1/2}]^2$ & ${}^*1 \times 10^{-4} [V_{1/2}]^2$ \\
\hline
\end{tabular}
\caption[Pseudoscalar leptoquark limits]{Updated pseudoscalar leptoquark limits. The starred entries are previous limits from reference \cite{Davidson:1993qk} which we believe to be erroneous, as they seem to have been obtained by treating incoherent contributions as being coherent. The daggered entries are previous limits from reference \cite{Herz:2002gq} which we believe to be erroneous as well, as they seem to have been obtained under the assumption that the interference from amplitudes that add coherently is always constructive. We use $[lq]$ to represent $\frac{m_{lq}}{100\ \mathrm{GeV}}$.}
\label{tab:pslq}
\end{center}
\end{table}

The amplitudes for the axial-vector leptoquark contributions have the same structure as the Standard Model amplitude, since the effective interactions that lead to these contributions are
\begin{eqnarray}
\mathcal{L}_{\mathrm{eff}} &\supset& \frac{\lambda^{ij}_{LS_0}\lambda^{kl*}_{LS_0}}{2m_{S_0}^2}(\bar{d}_L^{j,c}\gamma^{\mu}u_L^{l,c})(\bar{e}_L^{k}\gamma_{\mu}\nu_L^{i}) \\
&-& \frac{\lambda^{ij}_{LS_1}\lambda^{kl*}_{LS_1}}{2m_{S_1}^2}(\bar{d}_L^{j,c}\gamma^{\mu}u_L^{l,c})(\bar{e}_L^{k}\gamma_{\mu}\nu_L^{i}) \\
&+& \frac{\lambda^{ij}_{LV_0}\lambda^{kl*}_{LV_0}}{m_{V_0}^2}(\bar{u}_L^j\gamma^{\mu}d_L^l)(\bar{e}_L^{k}\gamma_{\mu}\nu_L^{i}) \\
&-& \frac{\lambda^{ij}_{LV_1}\lambda^{kl*}_{LV_1}}{m_{V_1}^2}(\bar{u}_L^j\gamma^{\mu}d_L^l)(\bar{e}_L^{k}\gamma_{\mu}\nu_L^{i}) +~\mathrm{h.c.}
\end{eqnarray}
Since these contributions are chirally suppressed, they will not be bounded as strongly as their pseudoscalar counterparts. They have the effect of changing the ratio of branching ratios to~\cite{Davidson:1993qk}
\begin{eqnarray}
R_\pi &\approx& R_\pi\mathrm{(SM)} \left(1 + \left(\frac{1}{2}\right) \frac{|\lambda^{11}_{LS}|^2}{\sqrt{2}G_{F}m^2} -  \left(\frac{1}{2}\right) \frac{|\lambda^{21}_{LS}|^2}{\sqrt{2}G_{F}m^2} \right) \\
R_\pi &\approx& R_\pi\mathrm{(SM)} \left(1 + \frac{|\lambda^{11}_{LV}|^2}{\sqrt{2}G_{F}m^2} - \frac{|\lambda^{21}_{LV}|^2}{\sqrt{2}G_{F}m^2} \right)
\end{eqnarray}
Again we limit only one product of couplings at a time, resulting in the bounds that are displayed in table \ref{tab:avlq}.
\begin{table}
\begin{center}
\begin{tabular}{|c|c|c|}
\hline
Product of couplings & New limit & Previous limit from \cite{Davidson:1993qk} or \cite{Herz:2002gq} \\
\hline
$|\lambda^{11}_{LS_0}|^2$ & $3.6 \times 10^{-4} [S_0]^2$ & $4 \times 10^{-3} [S_0]^2$ \\
$|\lambda^{21}_{LS_0}|^2$ & $1.7 \times 10^{-3} [S_0]^2$ & $4 \times 10^{-3} [S_0]^2$ \\
\hline
$|\lambda^{11}_{LS_1}|^2$ & $3.6 \times 10^{-4} [S_1]^2$ & $4 \times 10^{-3} [S_1]^2$ \\
$|\lambda^{21}_{LS_1}|^2$ & $1.7 \times 10^{-3} [S_1]^2$ & $4 \times 10^{-3} [S_1]^2$ \\
\hline
$|\lambda^{11}_{LV_0}|^2$ & $1.8 \times 10^{-4} [V_0]^2$ & $3.4 \times 10^{-4} [V_0]^2$ \\
$|\lambda^{21}_{LV_0}|^2$ & $8.7 \times 10^{-4} [V_0]^2$ & $1.7 \times 10^{-3} [V_0]^2$ \\
\hline
$|\lambda^{11}_{LV_1}|^2$ & $1.8 \times 10^{-4} [V_1]^2$ & $1.7 \times 10^{-3} [V_1]^2$ \\
$|\lambda^{21}_{LV_1}|^2$ & $8.7 \times 10^{-4} [V_1]^2$ & $3.4 \times 10^{-4} [V_1]^2$ \\
\hline
\end{tabular}
\caption[Axial-vector leptoquark limits]{Updated axial-vector leptoquark limits. We use $[lq]$ to represent $\frac{m_{lq}}{100\ \mathrm{GeV}}$.}
\label{tab:avlq}
\end{center}
\end{table}

%%%%%%%%%%%%%%%%%%%%%%%%%%%%%%%%%%%%%%%%%%%%%%%%%%%%%%%%%%%%%%%%%%%%%%
%%%%%%%%%%%%%%%%%%%%%%%%%%%%%%%%%%%%%%%%%%%%%%%%%%%%%%%%%%%%%%%%%%%%%%

\section{Extra Higgs Bosons}

Many extensions of the Standard Model incorporate extra Higgs particles, which may contribute to pion decay \cite{Donoghue:1978cj}. We consider the general case of two Higgs doublets with hypercharge $+1$,
\begin{equation}
\phi_i = \begin{pmatrix} \phi_i^+ \\ \phi_i^0 \end{pmatrix}
\end{equation}
for $i = 1,2$. In general, both of the Higgs doublets may acquire vacuum expectation values (vevs) in their lower components (to preserve electric charge) after spontaneous symmetry breaking. By appropriate $U(1)$ rotations, these vevs may be chosen to be real. Now, by a rotation in $\phi_1$-$\phi_2$ space, we may select out a particular combination of the Higgs doublets, $H$, that has a vanishing vev. (The orthogonal combination, with a non-trivial vev, assumes the role of the Standard Model Higgs.) The couplings of $H$ to the quarks and leptons is fixed only by gauge invariance, and in general we may have
\begin{equation}
\mathcal{L} \supset A^{ij} (\bar{Q}_L^i H) D_R^j + B^{ij} (\bar{Q}_L^i H^*) U_R^j + C^{ij} (\bar{L}_L^i H) E_R^j +~\mathrm{h.c.}
\end{equation}
in terms of the chiral quark and lepton fields, where $A$, $B$, and $C$ are arbitrary $3 \times 3$ matrices, and parentheses indicate $SU(2)$ singlets. 

Now, the coupling matrices may have significant off-diagonal terms, which would lead to flavour-changing neutral currents. In addition, the couplings are complex in general, leading to CP-violating interactions. Models that incorporate extra Higgs doublets thus usually impose consraints on their allowed couplings. For instance, in the Minimal Supersymmetric Standard Model, one Higgs doublet couples to the up-type quarks, while the other doublet couples to the down-type quarks and leptons. Such a scheme eliminates the possibility of flavour-changing neutral currents at tree level. However, these constraints also remove the contribution of the Higgs sector to pion decay, and so we place no \emph{a priori} restrictions on the $H$ couplings here.

The parts of the above Lagrangian terms that contribute to pion decay are
\begin{equation}
\mathcal{L} \supset A^{11} \bar{u}_L^1 H^+ d_R^1 + B^{11} \bar{d}_L^1 H^- u_R^1 + C^{11} \bar{\nu}_L^1 H^+ e_R^1 + C^{22} \bar{\nu}_L^2 H^+ e_R^2 +~\mathrm{h.c.}
\end{equation}
In terms of Dirac spinors, we have
\begin{equation}
\mathcal{L} \supset H^+ (A^{11} - B^{11*}) \bar{u} \gamma^5 d + H^- \left( \frac{C^{11*}}{2} \bar{e} (1 - \gamma^5) \nu_e + \frac{C^{22*}}{2} \bar{\mu} (1 - \gamma^5) \nu_\mu \right) +~\mathrm{h.c.},
\end{equation}
and it follows that
\begin{eqnarray}
\frac{\rho_e}{\Lambda^2} &=& \frac{\mathrm{Re}((A^{11} - B^{11*}) C^{11*})}{m_H^2}, \\
\frac{\rho_\mu}{\Lambda^2} &=& \frac{\mathrm{Re}((A^{11} - B^{11*}) C^{22*})}{m_H^2},
\end{eqnarray}
where $m_H$ is the mass of the $H$ field.

Assuming that only one of $A^{11}$ and $B^{11}$ is non-zero (in analogy with the doubling coupling dominance hypothesis), we obtain the limits shown in table \ref{tab:higgs}.

\begin{table}
\begin{center}
\begin{tabular}{|c|c|}
\hline
Product of couplings & Limit\\
\hline
$\mathrm{Re}(A^{11} C^{11*})$ & $1.2 \times 10^{-7} [H]^2$ \\
$\mathrm{Re}(B^{11*} C^{11*})$ & $1.2 \times 10^{-7} [H]^2$ \\
$\mathrm{Re}(A^{11} C^{22*})$ & $2.4 \times 10^{-5} [H]^2$ \\
$\mathrm{Re}(B^{11*} C^{22*})$ & $2.4 \times 10^{-5} [H]^2$ \\
\hline
\end{tabular}
\caption[Extra Higgs limits]{Limits on couplings of an extra Higgs doublet. We use $[H]$ to represent $\frac{m_H}{100\ \mathrm{GeV}}$.}
\label{tab:higgs}
\end{center}
\end{table}

%%%%%%%%%%%%%%%%%%%%%%%%%%%%%%%%%%%%%%%%%%%%%%%%%%%%%%%%%%%%%%%%%%%%%%%%
%%%%%%%%%%%%%%%%%%%%%%%%%%%%%%%%%%%%%%%%%%%%%%%%%%%%%%%%%%%%%%%%%%%%%%%%

\section{(Super)Compositeness}

%%%%%%%%%%%%%%%%%%%%%%%%%%%%%%%%%%%%%%%%%%%%%%%%%%%%%%%%%%%%%%%%%%%%%%%%

\subsection{Compositeness}
Models of composite quarks and leptons in general induce four-fermion contact interactions, suppressed by two powers of the compositeness scale. In general, these interactions may yield scalar, pseudoscalar, vector, axial-vector, or tensor contributions to Standard Model amplitudes. Ignoring tensor interactions, note that for two Dirac spinors $\psi^1$ and $\psi^2$,
\begin{align}
 \bar{\psi}^1 \psi^2 &= \bar{\psi}^1_L \psi^2_R + \bar{\psi}^1_R \psi^2_L \\
 \bar{\psi}^1 \gamma^5 \psi^2 &= \bar{\psi}^1_L \psi^2_R - \bar{\psi}^1_R \psi^2_L \\
 \bar{\psi}^1 \gamma^\mu \psi^2 &= \bar{\psi}^1_R \gamma^\mu \psi^2_R + \bar{\psi}^1_L \gamma^\mu \psi^2_L \\
 \bar{\psi}^1 \gamma^\mu \gamma^5 \psi^2 &= \bar{\psi}^1_R \gamma^\mu \psi^2_R - \bar{\psi}^1_L \gamma^\mu \psi^2_L
\end{align}
In particular, vector and axial-vector interactions couple spinors of the same chirality, while scalar and pseudoscalar interactions ``flip'' chirality. All gauge boson-fermion couplings in the Standard Model are combinations of vector and axial-vector interactions, so it is often useful to consider chirality-conserving structures, since such contributions interfere with vector boson exchange. The Particle Data Group convention for the general form of these interactions is \cite{Yao:2006px,Eichten:1983hw}
\begin{equation}
 \mathcal{L} \supset \frac{g^2}{2\Lambda^2} (\eta_{LL} \bar{\psi}_L \gamma^\mu \psi_L \bar{\psi}_L \gamma_\mu \psi_L + \eta_{RR} \bar{\psi}_R \gamma^\mu \psi_R \bar{\psi}_R \gamma_\mu \psi_R + 2 \eta_{LR} \bar{\psi}_L \gamma^\mu \psi_L \bar{\psi}_R \gamma_\mu \psi_R),
\end{equation}
where $g$ is an arbitrary coupling, and the $\psi$ are fermionic fields that may be either leptons or quarks. Usually, conservation of baryon number and lepton number is assumed, so that only operators with four leptons, four quarks, or two leptons and two quarks are considered \cite{Yao:2006px}.  Following convention, we will choose $\frac{g^2}{4\pi}=1$ and set each $\eta$ to either $0$ or $\pm 1$, then extract a limit on $\Lambda$.

We consider the effect of the operator
\begin{equation}
\frac{g^2}{2\Lambda^2} (\bar{u}_L \gamma^\mu d_L) (\bar{e}_L \gamma^\mu (\nu_e)_L),
\end{equation}
with $\frac{g^2}{4\pi}=1$ chosen as described above, and limit the compositeness scale $\Lambda$. Although its contribution is chirally suppressed (as expected, since the operator conserves chirality), the contact operator interferes coherently with the Standard Model contribution to pion decay. It changes the ratio of branching ratios to
\begin{equation}
R_\pi \approx R_\pi\mathrm{(SM)} \left(1 + \frac{\sqrt{2}\pi}{G_F \Lambda^2}\right),
\end{equation}
leading to the bound
\begin{equation}
\Lambda > 8.5\ \mathrm{TeV}.
\end{equation}

In the massless fermion limit, only vector and axial-vector operators, not scalar and pseudoscalar operators, can interfere with vector boson exchange, since $\psi_L$ and $\psi_R$ decouple from each other as $m \rightarrow 0$. Now, although the massless limit is a good approximation at high-energy colliders, pion decay \emph{requires} a chirality flip (and hence a mass term), and so (pseudo)scalar operators may be tightly constrained using the pion branching ratios. We thus place limits on the pseudoscalar interactions
\begin{equation}
\frac{g^2}{2\Lambda^2} (\bar{u} \gamma^5 d) (\bar{e} \gamma^5 \nu),
\end{equation}
again assuming $\frac{g^2}{4\pi}=1$, and neglecting the analogous operator which contributes to the muon decay mode. For the operator involving the electron neutrino, the effective strength of the coupling is
\begin{equation}
\frac{\rho_e}{\Lambda^2} = \frac{\pi}{\Lambda^2},
\end{equation}
and so we obtain a limit of
\begin{equation}
\Lambda > 5.2 \cdot 10^2~\mathrm{TeV}.
\end{equation}
For the operators involving the muon and tau neutrinos, we have
\begin{equation}
 \frac{\rho'_e}{\Lambda^2} = \frac{\pi}{\Lambda^2},
\end{equation}
and the limit is
\begin{equation}
 \Lambda > 1.0 \cdot 10^2~\mathrm{TeV}.
\end{equation}
As expected, these limits are much better than that on the vector operator above, because pseudoscalar contributions to pion decay are not chirally suppressed.

Finally, we limit the ($\bar{q}q$) scalar contact interactions, which hypothetically might arise in preon models where the constituent binding on preons that form quarks was effectively vectorlike, giving both chiral states of quarks and parity conserving scalar quark bilinear contact interactions, without their parity odd pseudoscalar counterparts. These would appear in the effective dimension six operators
\begin{equation}
 \frac{g^2}{2\Lambda^2} (\bar{u} d) (\bar{e} \nu),
\end{equation}
which are not chirally suppressed. Although these operators do not contribute to pion decay at tree level, because the pion is a pseudoscalar particle, renormalization leads to a modification of $R_\pi$ at one loop. We may see this phenomenon explicitly by working in the quark chiral basis, writing
\begin{equation}
 \frac{g^2}{2\Lambda^2} (\bar{u} d) (\bar{e} \nu) = \frac{g^2}{2\Lambda^2} (\bar{u}_R d_L) (\bar{e}_R \nu_L) + \frac{g^2}{2\Lambda^2} (\bar{u}_L d_R) (\bar{e}_R \nu_L),
\end{equation}
where we have ignored right-handed neutrinos. The two chiral operators above are renormalized differently, so if they start out with equal coefficients at the scale $\Lambda$, then their coefficients will be different at the weak scale, where the renormalization group equations change as we integrate out the $W$. The difference between the coefficients corresponds to a pseudoscalar operator, which may then be limited as before. We write the operators
\begin{align}
 \mathcal{O}_1 &= c_1 (\bar{e}_R L_L) (\bar{Q}_L d_R) \nonumber \\
 \mathcal{O}_2 &= c_2 (\bar{e}_R L_L) (\bar{u}_R Q_L) \nonumber \\
 \mathcal{O}_3 &= c_3 \left(-\frac{1}{8}\right) (\bar{e}_R \sigma^{\mu\nu} L_L) (\bar{u}_R \sigma_{\mu\nu} Q_L),
\end{align}
where we have suppressed $SU(2)$ indices and taken $\sigma^{\mu\nu} = \frac{i}{2} [\gamma^\mu, \gamma^\nu]$. The coefficients satisfy the partially coupled renormalization group equations
\begin{equation}
 M \frac{\partial c_i}{\partial M} = \frac{1}{32\pi^2} \gamma^{ij} c_j,
\end{equation}
where summation over repeated indices is implied and the anomalous dimension matrix is~\cite{Campbell:2003ir}
\begin{equation}
 \gamma = \begin{pmatrix} 6 g^2 + \frac{98}{9} g^{'2} & 0 & 0 \\ 0 & 6 g^2 + \frac{128}{9} g^{'2} & 6 g^2 + 10 g^{'2} \\ 0 & \frac{9}{2} g^2 + \frac{15}{2} g^{'2} & 12 g^2 + \frac{103}{9} g^{'2} \end{pmatrix}.
\end{equation}
In this expression, $g$ and $g'$ are the $SU(2)$ and $U(1)$ couplings, respectively, and are renormalized according to~\cite{Ramond:1999vh}
\begin{align}
 M \frac{\partial g}{\partial M} &= \frac{1}{32\pi^2} \left(-\frac{19}{3}\right) g^3, \\
 M \frac{\partial g'}{\partial M} &= \frac{1}{32\pi^2} \left(\frac{41}{5}\right) g^{'3}.
\end{align}
Starting with $c_1 = c_2 = \frac{2\pi}{\Lambda^2}$, $c_3 = 0$ at $M = 1~\mathrm{TeV}$, numerical integration yields $|c_2 - c_1| \approx (2.5 \cdot 10^{-3}) \frac{2\pi}{\Lambda^2}$ at $M = 100~\mathrm{GeV}$, which is approximately the weak scale. We thus effectively have
\begin{align}
 \frac{\rho_e}{\Lambda^2} &= (2.5 \cdot 10^{-3}) \frac{\pi}{\Lambda^2}, \\
 \frac{\rho'_e}{\Lambda^2} &= (2.5 \cdot 10^{-3}) \frac{\pi}{\Lambda^2},
\end{align}
which lead to the limits
\begin{align}
 \Lambda &> 26~\mathrm{TeV~(electron~neutrino)}, \\
 \Lambda &> 5.0~\mathrm{TeV~(other~neutrinos)}.
\end{align}
These limits are weaker than those in the pseudoscalar case, because although the scalar contributions to pion decay are not chirally suppressed, they involve one loop diagrams (through the renormalization group equations), which are suppressed by a power of the gauge coupling relative to their tree level counterparts.

%%%%%%%%%%%%%%%%%%%%%%%%%%%%%%%%%%%%%%%%%%%%%%%%%%%%%%%%%%%%%%%%%%%%%%%%%%

\subsection{Supercompositeness}
\label{sect:supercomp}
\begin{figure}
\begin{center}
\includegraphics[height=0.5\textheight]{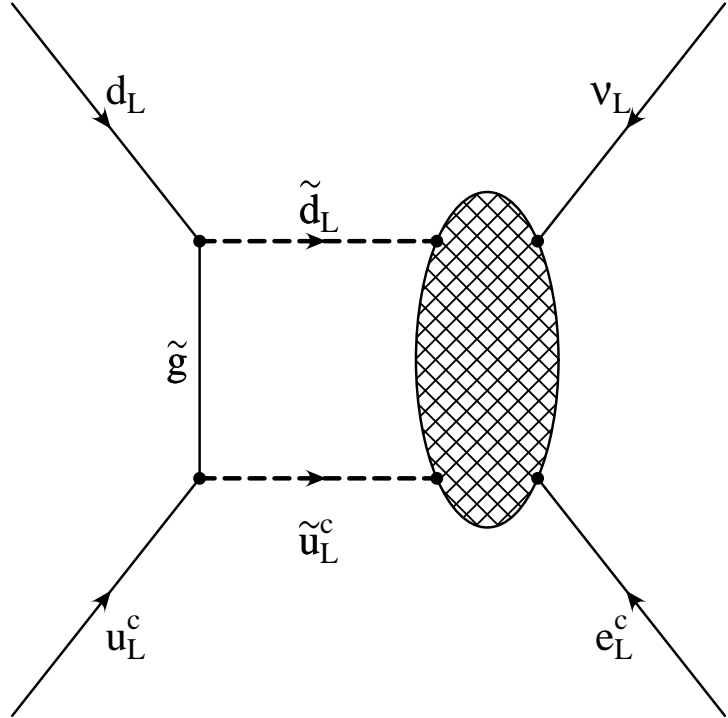}
\caption[Supercompositeness contribution to pion decay]{A contribution to pion decay involving gluino dressing of a dimension-five supercompositeness operator.}
\label{fig:supercomp}
\end{center}
\end{figure}
Supercomposite models treat (s)quarks and (s)leptons as composite particles, with the compositeness scale above the scale of supersymmetry breaking so that the induced contact operators include supersymmetric particles. The leading contributions of such models to pion decay come from dimension-five operators in the Lagrangian. (The operators are dimension-five because they involve two scalars and two fermions, as opposed to the dimension-six compositeness operators of the previous section, which involved four fermions.) These operators are suppressed by one power of the supercompositeness scale, and come from two types of terms in the Lagrangian. The F-terms come from the superpotential, as discussed in section \ref{sect:rpvsusy}, while the D-terms are the Lagrangian terms that can not be obtained from the superpotential, because they involve the complex conjugates of superfields. The terms which respect the $SU(3) \times SU(2) \times U(1)$ gauge symmetries and R-parity are~\cite{Weinberg:1981wj}
\begin{align}
 &(L_L E_L^c H_u^*)_D, (Q_L D_L^c H_u^*)_D, (Q_L U_L^c H_d^*)_D, (L_L L_L H_u H_u)_F, \nonumber \\
 &(Q_L Q_L U_L^c D_L^c)_F, (Q_L U_L^c L_L E_L^c)_F, (Q_L Q_L Q_L L_L)_F, (U_L^c U_L^c D_L^c E_L^c)_F.
\end{align}
The contributions to pion decay from the D-terms involve the couplings of Higgs particles to first-generation quarks and leptons, which are small, and so we only consider the F-term contributions to pion decay. Examining the above operators, we see that there is only one F-term that contributes to pion decay, $\frac{1}{M} Q_L U_L^c L_L E_L^c$, where $M$ is the supercompositeness scale. We may calculate its contribution by supersymmetric loop dressing of the dimension-five compositeness operator \cite{Campbell:1987ep}. The dominant contribution to the amplitude will be from gluino exchange, shown in figure \ref{fig:supercomp}. After performing the loop integral, taking $m_g \sim m_{\tilde{u}} \sim m_{\tilde{d}}$, we find that the amplitude associated with this diagram is
\begin{equation}
\mathcal{M} = \frac{\alpha_s}{2\sqrt{6}\pi}\frac{1}{M m_g} \left[\bar{l}\left(1-\gamma^5\right)\nu_l\right] \tilde{f_{\pi}},
\end{equation}
up to a complex phase, where $M$ is the supercompositeness scale and $m_g$ is the gluino mass. Neglecting the effect of the contribution to the muon decay channel, we may equate 
\begin{equation}
\frac{\rho_e}{\Lambda^2} = \frac{\alpha_s}{2\sqrt{6}\pi}\frac{1}{M m_g}.
\end{equation}
For $\alpha_s \sim 0.1$, this leads to the limit
\begin{equation}
M m_g > 6 \cdot 10^2\ \mathrm{TeV}^2.
\end{equation}
Taking $m_g \sim 1\ \mathrm{TeV}$ for illustration, we are left with a supercompositeness bound of
\begin{equation}
M > 6 \cdot 10^2\ \mathrm{TeV}.
\end{equation}

%%%%%%%%%%%%%%%%%%%%%%%%%%%%%%%%%%%%%%%%%%%%%%%%%%%%%%%%%%%%%%%%%%%%%%%%
%%%%%%%%%%%%%%%%%%%%%%%%%%%%%%%%%%%%%%%%%%%%%%%%%%%%%%%%%%%%%%%%%%%%%%%%

\section{Strong Gravity}

Recently there has been interest in models which address
the hierarchy problem by invoking extra dimensions of spacetime~\cite{ArkaniHamed:1998rs,Randall:1999ee} (for a review see \cite{Hewett:2002hv}). 
In these models gravity freely propagates in the extra dimensions,  
while normal standard model fields are either confined to a $3 + 1$ 
dimensional brane, or have otherwise localized wavefunctions in the 
extra dimensions.
%%%%%%%%%%%%%%%%%%%%%%%%%%%%%%%%%%%%%%%%%%%%%%%%%%%%%%%%%%%%%%%%%%%%%%%%
%There have been two general strategies proposed to generate the large 
%apparent hierarchy between the four-dimensional Planck scale and the 
%electroweak scale. In the first strategy gravity propagates in extra 
%flat dimensions beyond the four spacetime dimensions to which matter 
%is confined. The resulting Gauss-law dilution of the gravitational 
%force means that at the scale of the higher-dimensional Planck mass 
%is related to the scale of the effective large-distance four-dimensional 
%Planck mass by a factor that involves the volume of the large compactified %extra dimensions. 
%The second strategy invokes geometrically warped extra dimensions. In 
%models of this class the relative smallness of the electroweak scale
%relative to the Planck scale of gravity is understood to be a redshift 
%effect, due to the dynamical gravitational distortion of the spacetime.
%%%%%%%%%%%%%%%%%%%%%%%%%%%%%%%%%%%%%%%%%%%%%%%%%%%%%%%%%%%%%%%%%%%%%%%
The net result is that in these models the higher dimensional 
gravitational theory has strong gravitational interactions, with a 
fundamental Planck mass at the TeV scale, and the apparent weakness 
of four dimensional gravity is due to dilution or redshift 
effects.

%%%%%%%%%%%%%%%%%%%%%%%%%%%%%%%%%%%%%%%%%%%%%%%%%%%%%%%%%%%%%%%%%%%%%%%
Black holes will exist as states in the higher-dimensional theory with masses of order the higher-dimensional Planck scale and above. Collisions of standard model quanta at sufficiently high energies should then be able to produce higher-dimensional black holes at the location of the standard model brane \cite{Dimopoulos:2001hw, Giddings:2001bu, Meade:2007sz}. 
Also quanta will generically be non-local on the scale of the higher-dimensional Planck length; attempts to localize the quanta on shorter distance scales would involve super-Planckian momentum and energy transfer, resulting in the quantum formation a black hole of intrinsic size at least the Planck length. As such we would expect all quanta to have composite "overlap" interactions with a length scale set by the higher dimensional Planck length. At low energies, for quarks and leptons, the leading effect of these composite "contact" interactions would be described by four-fermion dimension-six operators, 
of the type analyzed in the previous section on quark/lepton 
compositeness; since the compositeness is due to (quantum) gravitational non-locality, the compositeness scale $\Lambda$ would be set by the higher-dimensional Planck mass, which is the dimensional parameter governing the Einstein action and which gives the scale at which non-perturbative gravitational effects become important.  

%%%%%%%%%%%%%%%%%%%%%%%%%%%%%%%%%%%%%%%%%%%%%%%%%%%%%%%%%%%%%%%%%%%%%%%%
So in the absence of wave-function effects corresponding to a separation 
of our colliding quanta in the extra dimension(s), we should expect both 
dimension-six compositeness contact interactions in the scattering of 
quanta with scale set by $M_{p}$, and the production of physical higher-dimensional black holes in the collisions of quanta provided the centre of mass energy exceeded $M_{p}$ to be above threshold for the production of a (quasi-classical) black hole.
Our results then have serious implications for the potential observability of black holes, or specific effects of non-perturbative gravitation, at upcoming colliders like the LHC. Already reference \cite{Meade:2007sz} has noted that the LEP limits on the operator ${\Lambda^{+}_{LL}}(eedd) > 26.4$ TeV (in the same PDG normalization that we have adopted) would preclude study of quantum gravity at the LHC unless some degree of wave-function suppression effects were invoked. The limits on the compositeness scale  (which for gravitationally induced compositeness should be universal and set by the Planck scale) that we have derived in the previous section are more than an order of magnitude more severe. For the non-supersymmetric case we were able to bound the compositeness scale as,
\begin{equation}
\Lambda > 5.2 \cdot 10^2~\mathrm{TeV}.
\end{equation}
while in the supersymmetric case the limit was even stronger.
This implies that if nature is described by an extra dimensional theory with strong higher-dimensional gravity, then either the Planck scale of the higher dimensional theory is more than an order of magnitude beyond the gravitational scales that will be accessible to direct study at the LHC, or else there are very large effects in the theory associated with the wave-function separation in the extra dimension(s). In turn this implies that studies of the observability and signatures of gravitational effects at the LHC, which are usually undertaken ignoring such wave function effects, are studies adapted to theoretical cases already excluded by leptonic pion decay; to realistically assess the signatures for extra dimensional strong gravity physics at the LHC new studies are required, which incorporate precisely the kind of wave-function suppressions necessary to prevent the model from already being excluded by pion data. 
%%%%%%%%%%%%%%%%%%%%%%%%%%%%%%%%%%%%%%%%%%%%%%%%%%%%%%%%%%%%%%%%%%%%%%%%%
%%%%%%%%%%%%%%%%%%%%%%%%%%%%%%%%%%%%%%%%%%%%%%%%%%%%%%%%%%%%%%%%%%%%%%%%%

\section{Conclusions}

As we have seen, the ratio of branching ratios in leptonic pion decay represents a very powerful test of theories of physics beyond the electroweak standard model. We have seen, in particular, that it provides 
stringent limits on R-parity violating interactions in supersymmetric theories, on theories with light (electroweak scale as opposed to GUT or Planck scale) leptoquark degrees of freedom, on non-minimal models of extra doublet Higgs bosons, on models in which the quarks and leptons are composite both without supersymmetry (and with or without parity conservation in the substructure dynamics) and with supersymmetry, and on models with strong TeV scale gravitational interactions. 
Moreover, at present the theoretical prediction for the ratio $R_{\pi}$ is a factor of forty more precise than the experimental determination. Experiments presently underway should improve the experimental precision by up to an order of magnitude. Through comparison of these future more precise measurements of $R_{\pi}$ with the existing high precision 
standard model predictions, these ongoing experiments have a potential window of discovery for physics beyond the standard model, which for certain classes of models extends to very large mass scales. Either we will see evidence for physics beyond the standard model in the next round of $\pi \rightarrow l \nu$ experiments, or else we will have even more powerful constraints on putative standard model extentions.

%%%%%%%%%%%%%%%%%%%%%%%%%%%%%%%%%%%%%%%%%%%%%%%%%%%%%%%%%%%%%%%%%%%%%%%%%
%%%%%%%%%%%%%%%%%%%%%%%%%%%%%%%%%%%%%%%%%%%%%%%%%%%%%%%%%%%%%%%%%%%%%%%%%

\section{Acknowledgements}

We are very grateful to David Maybury for discussions, constructive criticism, and assistance with calculations. We also wish to thank Heather Logan for her careful reading of the thesis upon which this paper is based. This work is supported in part by the Natural Sciences and Engineering Research Council of Canada.
%%%%%%%%%%%%%%%%%%%%%%%%%%%%%%%%%%%%%%%%%%%%%%%%%%%%%%%%%%%%%%%%%%%%%%%%%
%%%% AND I FOR ONE WELCOME THE ARRIVAL OF OUR NEW ALIEN OVERLORDS! %%%%%%
%%%%%%%%%%%%%%%%%%%%%%%%%%%%%%%%%%%%%%%%%%%%%%%%%%%%%%%%%%%%%%%%%%%%%%%%%

%%%%%%%%%%%%%%%%%%%%%%%%%%%%%%%%%%%%%%%%%%%%%%%%%%%%%%%%%%%%%%%%%%%%%%%%%
%%%%%%%%%%%%%%%%%%%%%%%%%%%%%%%%%%%%%%%%%%%%%%%%%%%%%%%%%%%%%%%%%%%%%%%%%

\bibliographystyle{unsrt}
\bibliography{references}

\end{document}